\documentclass[pdftex,twocolumn,epjc3]{svjour3}  

\RequirePackage[T1]{fontenc}

\smartqed

\RequirePackage{graphicx}
\RequirePackage{mathptmx}
\RequirePackage{flushend}
\RequirePackage[colorlinks,citecolor=blue,urlcolor=blue,linkcolor=blue]{hyperref}

\journalname{Eur. Phys. J. C}

\RequirePackage{graphicx}
\RequirePackage{amsmath} 
\RequirePackage{mathtools} 
\RequirePackage{booktabs}
\RequirePackage[tableposition=bottom]{caption}
\RequirePackage{hyperref} 
\RequirePackage{makecell}
\RequirePackage{nicefrac}
\RequirePackage{physics}
\RequirePackage{cite}
\RequirePackage[a4paper, total={6.5in, 8.5in}]{geometry}
\RequirePackage{xcolor}
\RequirePackage{amssymb}
\RequirePackage{multirow}
\RequirePackage{hepunits}
\RequirePackage{afterpage}
\RequirePackage[utf8]{inputenc}
\RequirePackage{enumitem}
\RequirePackage[normalem]{ulem}
\RequirePackage{cuted}

\newcommand{\eg}{\textit{e.g.}\xspace}
\newcommand{\ie}{\textit{i.e.}\xspace}
\newcommand{\cf}{\textit{cf.}\xspace}

\def\figWidth{\linewidth}

\newcounter{comment}

\begin{document}

\title{Data-driven study of timelike Compton scattering}

\author{
O.~Grocholski\thanksref{e1,addr1}
\and
H.~Moutarde\thanksref{e2,addr2}
\and
B.~Pire\thanksref{e3,addr3}
\and
P.~Sznajder\thanksref{e4,addr4} 
\and
J.~Wagner\thanksref{e5,addr4}
}

\institute{
Institute of Theoretical Physics, Faculty of Physics, University of Warsaw, Pasteura 5, 02-093 Warsaw, Poland  \label{addr1}
\and 
~IRFU, CEA, Universit\'e Paris-Saclay, F-91191 Gif-sur-Yvette, France \label{addr2} 
\and 
~CPHT, CNRS, \'Ecole Polytechnique, I. P. Paris, F-91128 Palaiseau, France \label{addr3} 
\and 
~National Centre for Nuclear Research (NCBJ), Pasteura 7, 02-093 Warsaw, Poland \label{addr4}
}

\thankstext{e1}{e-mail: oskar.grocholski@gmail.com}
\thankstext{e2}{e-mail: herve.moutarde@cea.fr}
\thankstext{e3}{e-mail: bernard.pire@polytechnique.edu}
\thankstext{e4}{e-mail: pawel.sznajder@ncbj.gov.pl}
\thankstext{e5}{e-mail: jakub.wagner@ncbj.gov.pl}

\date{Received: date / Accepted: date}

\maketitle

\sloppy

%\allowdisplaybreaks

\begin{abstract}
In the framework of collinear QCD factorization, the leading twist scattering amplitudes for deeply virtual Compton scattering (DVCS) and timelike Compton scattering (TCS) are
 intimately related thanks to analytic properties of leading and next-to-leading order amplitudes. We exploit this welcome feature to make data-driven predictions for TCS observables to be measured in near future experiments. Using a recent extraction of DVCS Compton form factors from most of the existing experimental data for that process, we derive TCS amplitudes and calculate TCS observables only assuming leading-twist dominance. Artificial neural network techniques are used for an essential reduction of model dependency, while a careful propagation of experimental uncertainties is achieved with replica methods. Our analysis  allows for stringent tests of the leading twist dominance of DVCS and TCS amplitudes. Moreover, this study helps to understand quantitatively the complementarity of DVCS and TCS measurements to test the universality of generalized parton distributions, which is crucial \eg to perform the nucleon tomography.  
\end{abstract}

\section{Introduction}
\label{sec:introduction}

It is now widely recognized that generalized parton distributions (GPDs) \cite{Mueller:1998fv, Ji:1996nm, Radyushkin:1996nd} are among the best known tools used to explore internal structures of nucleons and nuclei in terms of partonic degrees of freedom: quarks and gluons. GPDs are studied in exclusive reactions in kinematical regimes that allow to apply QCD factorization theorems \cite{Mueller:1998fv, Radyushkin:1997ki, Ji:1998xh}. The formalism of GPDs provides a rigorous theoretical framework to study the 3D structure of hadrons \cite{Burkardt:2000za,Ralston:2001xs, Diehl:2002he} and it allows to reach the QCD energy-momentum tensor (see the reviews \cite{Diehl:2003ny, Belitsky:2005qn, Boffi:2007yc, Guidal:2013rya} and references therein). The latter can be used to evaluate the contribution of orbital angular momentum generated by partons and it helps to understand mechanical properties of partonic media, like energy density or radial and tangential pressures \cite{Polyakov:2018zvc, Lorce:2018egm}. 

Because of its relatively straightforward description and accessible cross section, deeply virtual Compton scattering (DVCS),
\begin{equation}
\gamma^*(q) N(p_1) \to \gamma(q') N'(p_2) \,,
\end{equation}
has been recognized to be the golden channel in GPD studies. Here, the symbols in the parentheses denote four-momenta of photons and nucleons. The collinear QCD factorization between GPDs and perturbatively calculable coefficient functions requires the photon virtuality, $Q^2=-q^2$, to be large, while the absolute value of Mandelstam variable $t=(p_2 - p_1)^2$ to be small, such as $-t/Q^2 \ll 1$.  DVCS was the first channel used to prove the usefulness of the GPD formalism in experiments where lepton beams were scattered off hadron targets, like Hall-A and CLAS at JLab, COMPASS at CERN, and HERMES, ZEUS and H1 at DESY. The extensive global effort to measure DVCS in various kinematic domains and for various combinations of charges and polarizations of beams and targets, gave so far about 30 observables collected over more than 2500 kinematic configurations. All these data, which were published over 17 years, were recently used in state-of-the-art global fits \cite{Moutarde:2018kwr, Moutarde:2019tqa} based on the open-source PARTONS framework \cite{Berthou:2015oaw}, which provides a homogeneous computational environment for all kind of GPD studies.

The crossed reaction to DVCS, timelike Compton scattering (TCS),
\begin{equation}
 \gamma(q) N(p_1) \to \gamma^*(q') N'(p_2) \,,   
\end{equation}
in the domain of large $Q'^2=+q'^2$ and nearly forward kinematics, is also a very promising process to probe GPDs \cite{Berger:2001xd,Pire:2011st,Moutarde:2013qs,Anikin:2017fwu}. This possibility has not yet been fully explored experimentally, but measurements at JLab are either under way \cite{JLAB_TCS_proposal} or planned \cite{JLAB_TCS_proposal2}. Detailed predictions have been presented within the leading order approximation \cite{Boer:2015fwa,Boer:2015cwa} for both proton and neutron targets. This process is also interesting in the case of ultraperipheral collisions at hadron colliders \cite{Pire:2008ea}, as nucleons and nuclei are intense sources of quasi-real photons.

Two other reactions that access the quark and gluon content of nucleons and nuclei with \emph{only} the electromagnetic probes, namely the double deeply virtual Compton scattering (DDVCS) \cite{Guidal:2002kt, Belitsky:2002tf},
\begin{equation}
 \gamma^*(q) N(p_1) \to \gamma^*(q') N'(p_2)\,,   
\end{equation}
with both $-q^2$ and $q'^2$ being large, and photoproduction of a photon pair\cite{Pedrak:2017cpp},
\begin{equation}
 \gamma(q) N(p_1) \to \gamma(q_1) \gamma(q_2) N'(p_2)\,,   
\end{equation}
with large invariant mass $M_{\gamma \gamma}^2=(q_1+q_2)^2$, are also very powerful processes to probe GPDs. However, they suffer from rather small cross sections that prevent any measurements at this moment. 

Amplitudes of deeply virtual meson production (DVMP),
\begin{equation}
\gamma^*(q) N(p_1) \to M(q') N'(p_2) \,,
\end{equation}
are known to obey the same factorization theorems \cite{Collins:1996fb} as DVCS and TCS amplitudes, and have been studied in great detail both experimentally and theoretically. Nowadays, the status of these analyses is quite mitigate, since most polarization tests of the validity of factorization (such as the dominance of the longitudinal virtual photon exchange in $\pi$, $\rho$ and $\omega$ electroproduction) are violated at moderate $Q^2$, where most of the available data exist (see the review \cite{Favart:2015umi} and references therein). 

DVCS and TCS are two independent sources of GPD information that can be extracted from experimental data. Additionally, the complementarity between both reactions observed up to the next-to-leading order coefficient functions is today the best known tool to test the validity of the collinear QCD factorization framework and the universality of GPDs. To reach this goal, we undertake to derive model independent predictions for TCS observables, assuming the current knowledge of DVCS amplitudes. In our previous paper \cite{Moutarde:2019tqa}, we have extracted DVCS amplitudes, \ie Compton form factors (CFFs), from a model independent global fit of data collected in various experiments. A direct extraction of amplitudes has allowed us in particular to not be limited by any order of QCD calculation of DVCS coefficient functions. 

We now extend our analysis of DVCS to the case of TCS. Due to the relation between DVCS and TCS amplitudes, which we derived in Ref.~\cite{Muller:2012yq} and which is based on the analyticity of next-to-leading order amplitudes in $Q^2$, we are able to make data-driven predictions for TCS observables. The only model dependence of those predictions comes from the assumption about the leading twist dominance and the restriction to LO and NLO in the strong coupling constant. Obtained results will be useful to check the universality of GPDs, but also to determine which TCS observables could be the best source of new information on CFFs and GPDs. The presented approach is a promising tool to be used in analyses of future TCS data.
\section{Relation between DVCS and TCS amplitudes}
\label{sec:DVCS_vs_TCS}
Both DVCS and TCS helicity amplitudes can be conveniently described in terms of Compton form factors (CFFs) \cite{Diehl:2003ny}:
\begin{gather}
^SM_{++++} = \sqrt{1-\xi^2}\left[^S{\cal H} +^S\widetilde{\cal H} -\frac{\xi^2}{1-\xi^2}(^S{\cal E}+^S\widetilde{\cal E})\right] \,, \nonumber \\
^SM_{-+-+} = \sqrt{1-\xi^2}\left[^S{\cal H} -^S\widetilde {\cal H} -\frac{\xi^2}{1-\xi^2}(^S{\cal E}-^S\widetilde{\cal E})\right] \,, \nonumber \\
^SM_{++-+} = \frac{\sqrt{t_0-t}}{2M}\left[^S{\cal E}-\xi ^S\widetilde{\cal E}\right] \,, \nonumber \\
^SM_{-+++} = -\frac{\sqrt{t_0-t}}{2M}\left[^S{\cal E}+\xi ^S\widetilde{\cal E}\right] \,,
\label{eq:CFF_S}
\end{gather}
and 
\begin{gather}
^TM_{+-+-} = \sqrt{1-\xi^2}\left[^T{\cal H} +^T\widetilde{\cal H} -\frac{\xi^2}{1-\xi^2}(^T{\cal E}+^T\widetilde{\cal E})\right] \,, \nonumber \\
^TM_{----} = \sqrt{1-\xi^2}\left[^T{\cal H} -^T\widetilde {\cal H} -\frac{\xi^2}{1-\xi^2}(^T{\cal E}-^T\widetilde{\cal E})\right] \,, \nonumber \\
^TM_{+---} = \frac{\sqrt{t_0-t}}{2M}\left[^T{\cal E}-\xi ^T\widetilde{\cal E}\right] \,, \nonumber \\
^TM_{--+-} = -\frac{\sqrt{t_0-t}}{2M}\left[^T{\cal E}+\xi ^T\widetilde{\cal E}\right] \,.
\label{eq:CFF_T}
\end{gather}
Here, $^X M_{\lambda' \mu' \lambda \mu}$ denotes the helicity amplitudes for DVCS ($X=S$) and TCS ($X=T$), $\lambda$ ($\lambda'$) is the helicity of the incoming (outgoing) proton, and $\mu$ ($\mu'$) is the helicity of the incoming (outgoing) photon. The CFFs ${\cal H},{\widetilde{\cal H}},{\cal E}$ and ${\widetilde{\cal E}}$ are functions of four variables: the square of four-momentum transfer $t = (p_2-p_1)^2$, the longitudinal momentum transfer (skewness) $\xi$, the photon virtuality $\mathcal{Q}$ and the factorization scale $\mu_F$. The latter is omitted in the following equations for the brevity. With $M$ standing for the mass of the nucleon, $t_0 = - 4 \xi^2 M^2 / (1-\xi^2)$ is the smallest absolute value of $t$ allowed at a fixed value of skewness (up to contributions suppressed by power corrections of the order of ${\cal Q}^2$). The variable $\tau = {Q'^2}/(2 p\cdot q)$ for TCS is the analog of the Bjorken variable $x_B = Q^2/(2 p\cdot q)$ for DVCS. The similar role played by these quantities reveals itself in their relations with $\xi$, which to the leading twist accuracy reads $\xi = \tau /(2 - \tau)$ for TCS and $\xi = x_B /(2 - x_B)$ for DVCS.

Factorization theorems allow to express CFFs in terms of perturbatively calculable coefficient functions $T^i$ and GPDs $F^i$, where $i = u, d, \ldots, g$ denotes a given parton type:
\begin{eqnarray}
{\cal F}(\xi, t,{\cal Q}^2) =
\int_{-1}^1\!\! dx\!
\sum_{i=u,d,\ldots, g}\!\! {T^i(x,\xi,{\cal Q}^2)} {F^i(x,\xi,t)} \,.
\label{eq:factorizedamplitude}
\end{eqnarray}
The coefficient functions $^S T$ for the spacelike case at LO and NLO read:
\begin{eqnarray}
{^ST^i} &\stackrel{\rm LO}{=}& {^SC_{0}^i} \\
 {^ST^i} &\stackrel{\rm NLO}{=}& {^SC_{0}^i} + \frac{\alpha_s(\mu_{R}^2)}{2\pi}\left[
 {^SC_1^i} + {^SC_{\mathrm{coll}}^i} \ln\frac{{\cal Q}^2}{\mu_{F}^2}\right] \,,
\label{eq:coefficients}
\end{eqnarray}
where $\mu_{R}$ is the renormalization scale. The expressions for ${^SC_0^i}$, ${^SC_1^i}$ and ${^SC_{\mathrm{coll}}^i}$ can be found in Ref. \cite{Muller:2012yq}.

Thanks to simple spacelike-to-timelike relations derived in Ref.~\cite{Muller:2012yq}, we can express the timelike coefficients by the spacelike ones in the following way:
\begin{eqnarray}
{^T}T^i &\stackrel{\rm LO}{=}& \pm {^S}T^{i\,*} \,,
\\
{^T}T^i &\stackrel{\rm NLO}{=}& \pm {^S}T^{i\,*} \mp i\pi \frac{\alpha_s(\mu_{R}^2)}{2\pi} {^S}C_{\mathrm{coll}}^{i\,*} \,, 
\label{fullrelation}
\end{eqnarray}
where upper (lower) sign is for (anti-)symmetric coefficient functions in $\xi$. For (anti-)symmetric CFFs $\cal H$ ($\widetilde{\cal H}$) this gives:
\begin{eqnarray}
{^T{\cal H}}  &\stackrel{\rm LO}{=}&  ^S{\cal H}^\ast \,,
\label{eq:cffH2cffTH_LO}
\\
{^T\widetilde{\cal H}}  &\stackrel{\rm LO}{=}&  -^S\widetilde{\cal H}^\ast \,,
\label{eq:cffH2cffTHt_LO}
\\
{^T{\cal H}}  &\stackrel{\rm NLO}{=}&  ^S{\cal H}^\ast - i\pi\, {\cal Q}^2\frac{\partial}{\partial {\cal Q}^2} {^S{\cal H}}^\ast \,,
\label{eq:cffH2cffTH}
\\
{^T\widetilde{\cal H}}  &\stackrel{\rm NLO}{=}&  -^S\widetilde{\cal H}^\ast + i\pi\, {\cal Q}^2\frac{\partial}{\partial {\cal Q}^2}  {^S\widetilde{\cal H}^\ast}\,.
\label{eq:cffHt2cffTHt}
\end{eqnarray}
The corresponding relations exist for (anti-)symmetric CFFs $\cal E$ ($\widetilde{\cal E}$).

In the recent study \cite{Moutarde:2019tqa}, the artificial neural network technique was employed to determine the spacelike CFFs from a global analysis of almost all DVCS measurements off a proton target. In this analysis the replica method was used to propagate experimental uncertainties to those of extracted quantities. Together with Eqs. (\ref{eq:cffH2cffTH_LO}-\ref{eq:cffHt2cffTHt}), this  creates an opportunity to perform model independent predictions for TCS, thus allowing for a quantitative assessment of the impact of the expected measurements.
    
\begin{figure}[!ht]
\includegraphics[width=\figWidth]{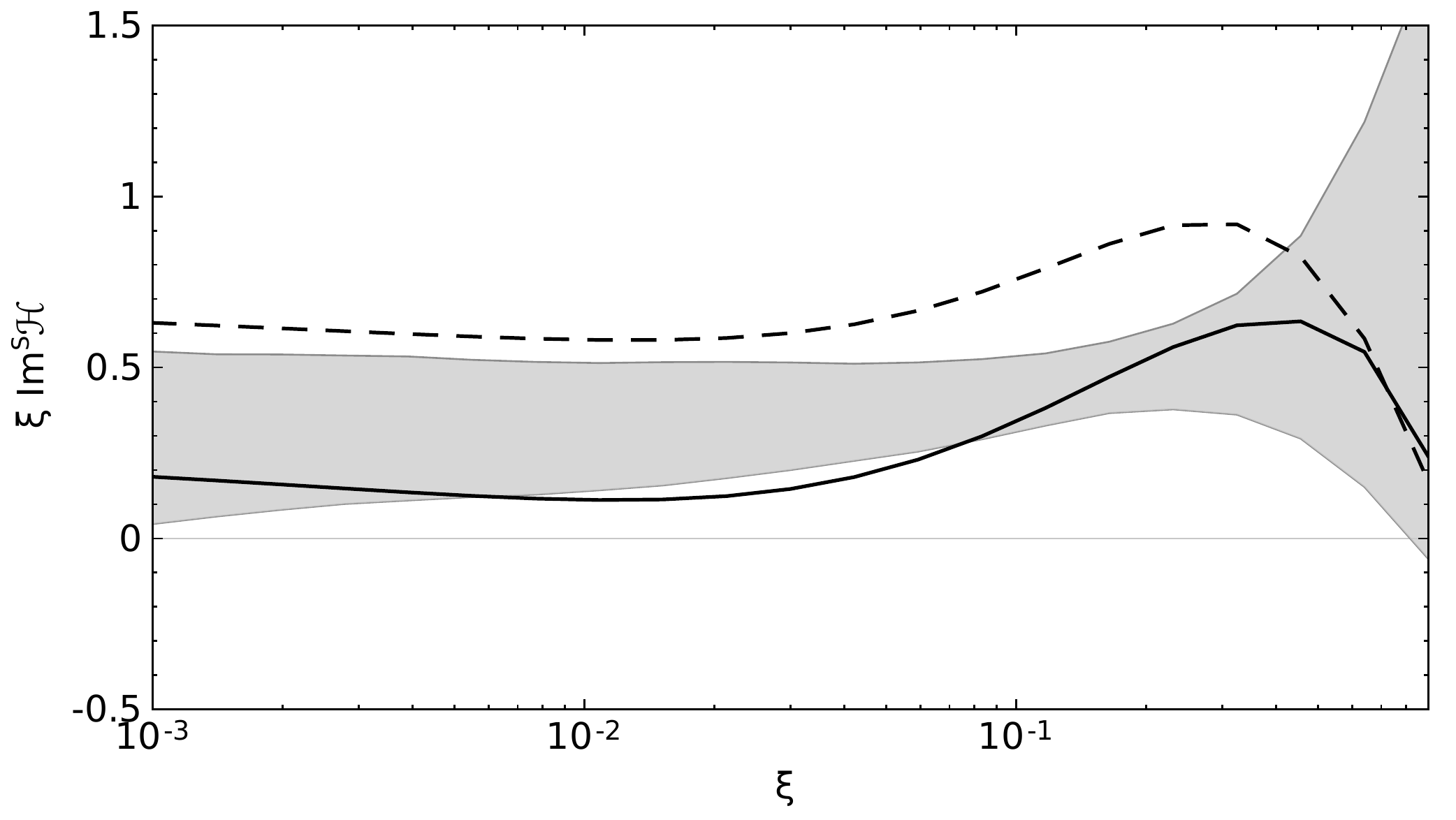}
\includegraphics[width=\figWidth]{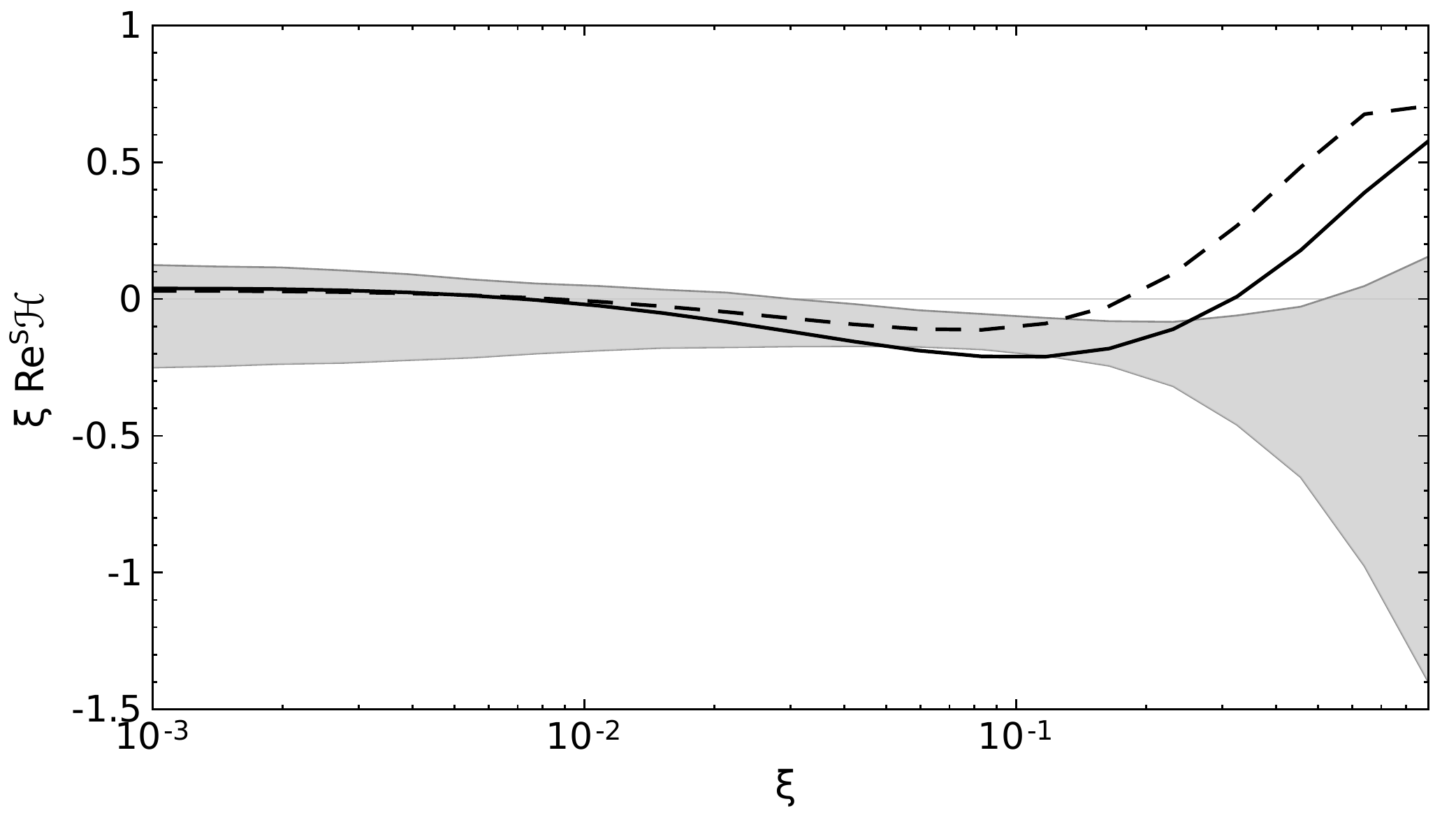}
\caption{Imaginary (up) and real (down) part of DVCS CFF $\xi ^S\mathcal{H}(\xi)$ for $Q^2 = 2$ GeV$^2$ and $t=-0.3$ GeV$^2$ as a function of $\xi$. The shaded gray bands correspond to the global fit of DVCS data presented in \cite{Moutarde:2019tqa} and they show 68\% confidence level for the uncertainties of presented quantities. The dashed (solid) lines correspond to the GK GPD model \cite{Goloskokov:2005sd, Goloskokov:2007nt, Goloskokov:2009ia} evaluated with LO (NLO) DVCS coefficient functions.}
\label{fig:DVCS_CFFs}
\end{figure}

\begin{figure}[!ht]
\begin{center}
\includegraphics[width=\figWidth]{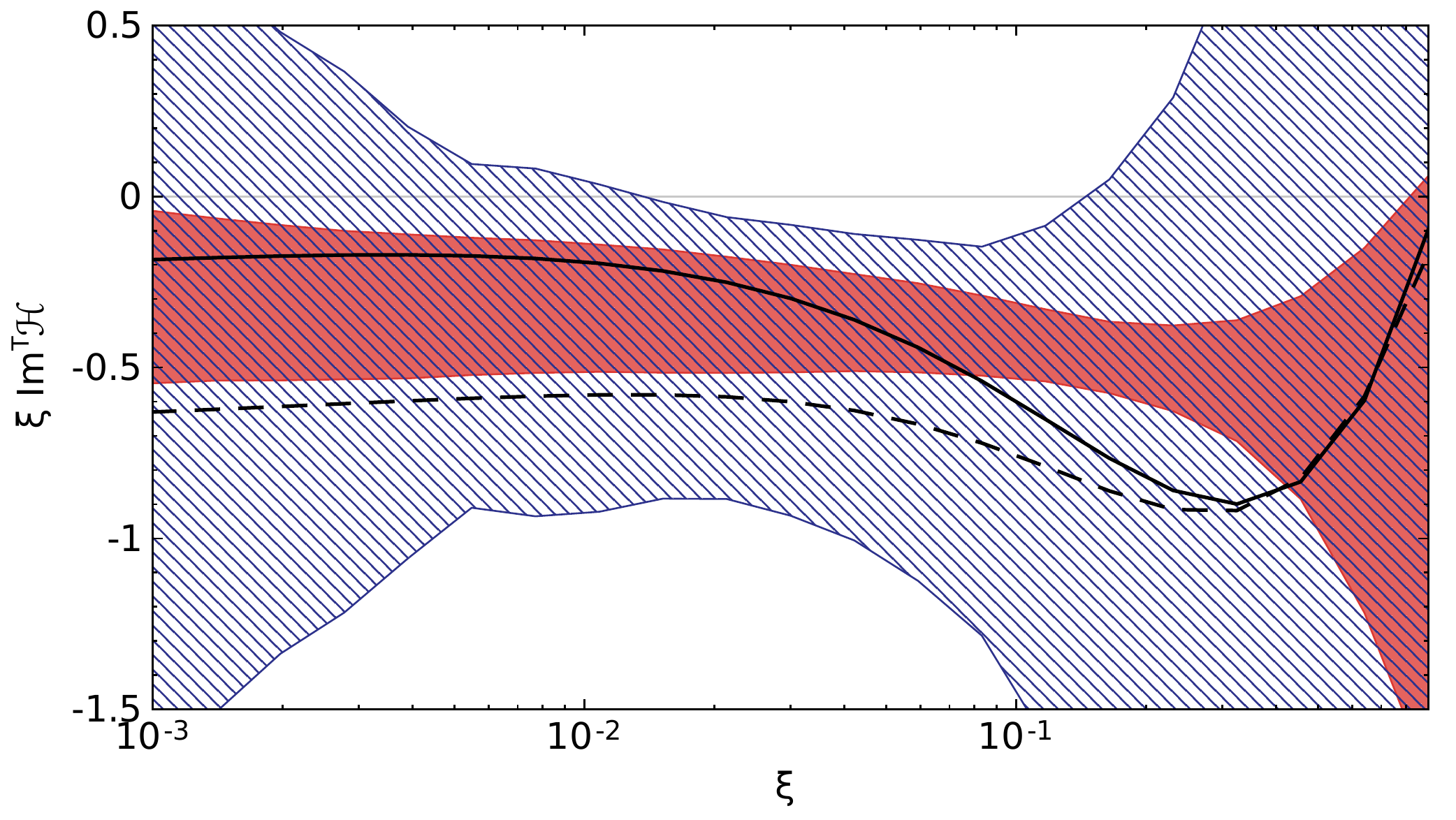}
\includegraphics[width=\figWidth]{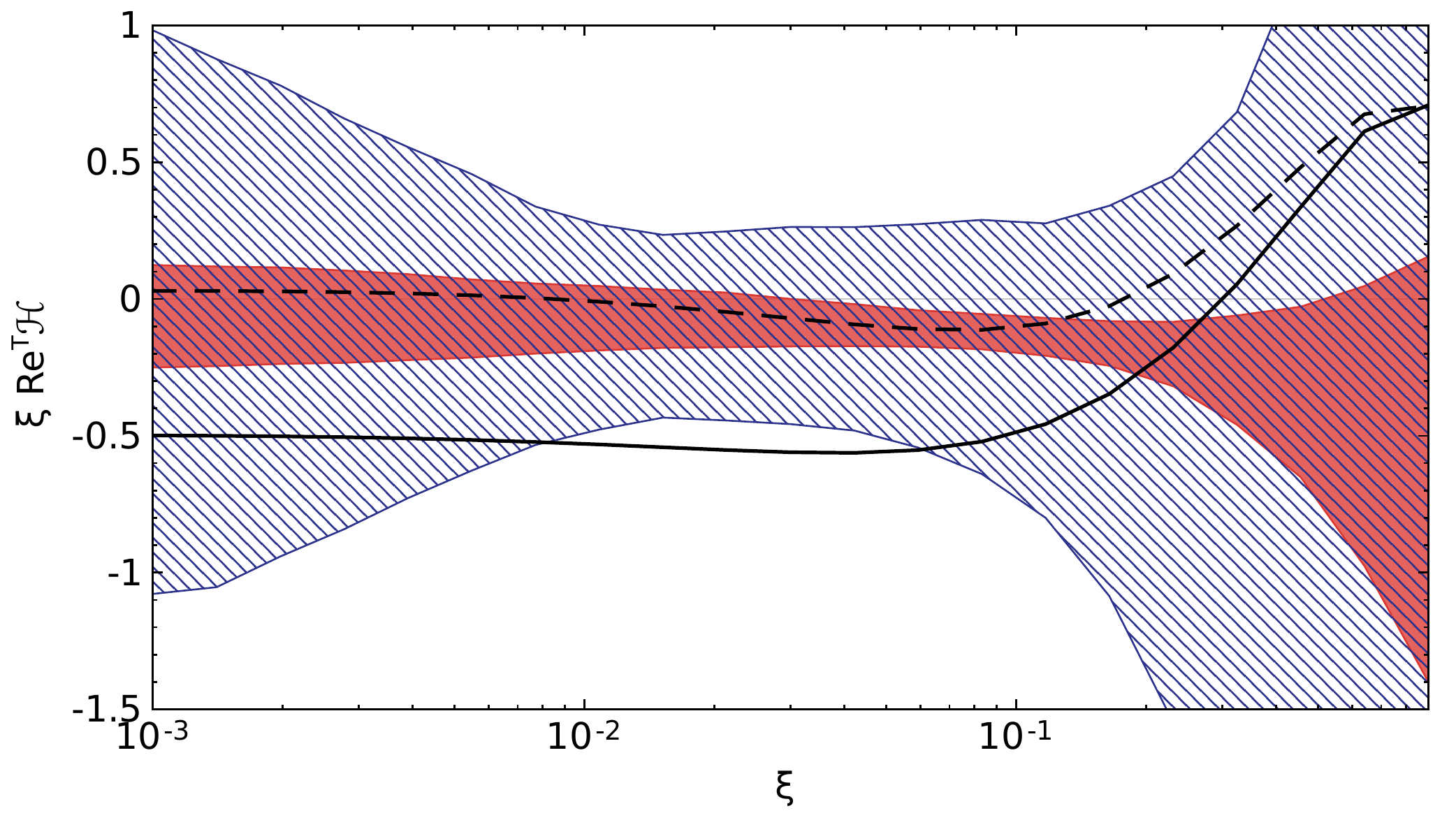}
\caption{Imaginary (up) and real (down) part of TCS CFF $\xi ^T\mathcal{H}(\xi)$ for $Q^2 = 2$ GeV$^2$ and $t=-0.3$ GeV$^2$ as a function of $\xi$. The shaded red (dashed blue) bands correspond to the data-driven predictions coming from the global fit of DVCS data presented in \cite{Moutarde:2019tqa} and they are evaluated using LO (NLO) spacelike-to-timelike relations. The bends show 68\% confidence level for the uncertainties of presented quantities. The dashed (solid) lines correspond to the GK GPD model \cite{Goloskokov:2005sd, Goloskokov:2007nt, Goloskokov:2009ia} evaluated with LO (NLO) TCS coefficient functions.}
\label{fig:TCS_CFFs}
\end{center}
\end{figure}

\begin{figure}[!ht]
\begin{center}
\includegraphics[width=\figWidth]{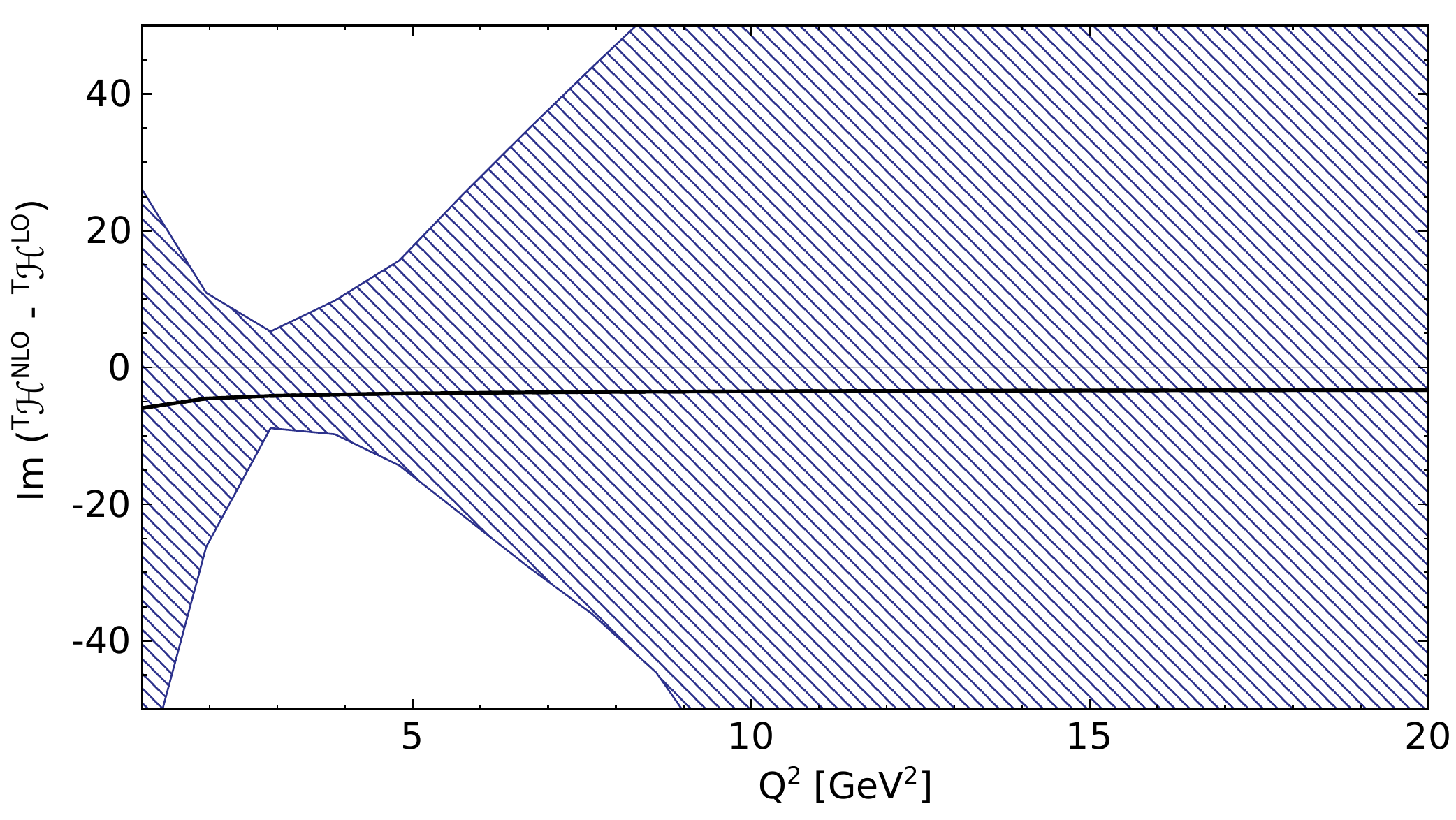}
\includegraphics[width=\figWidth]{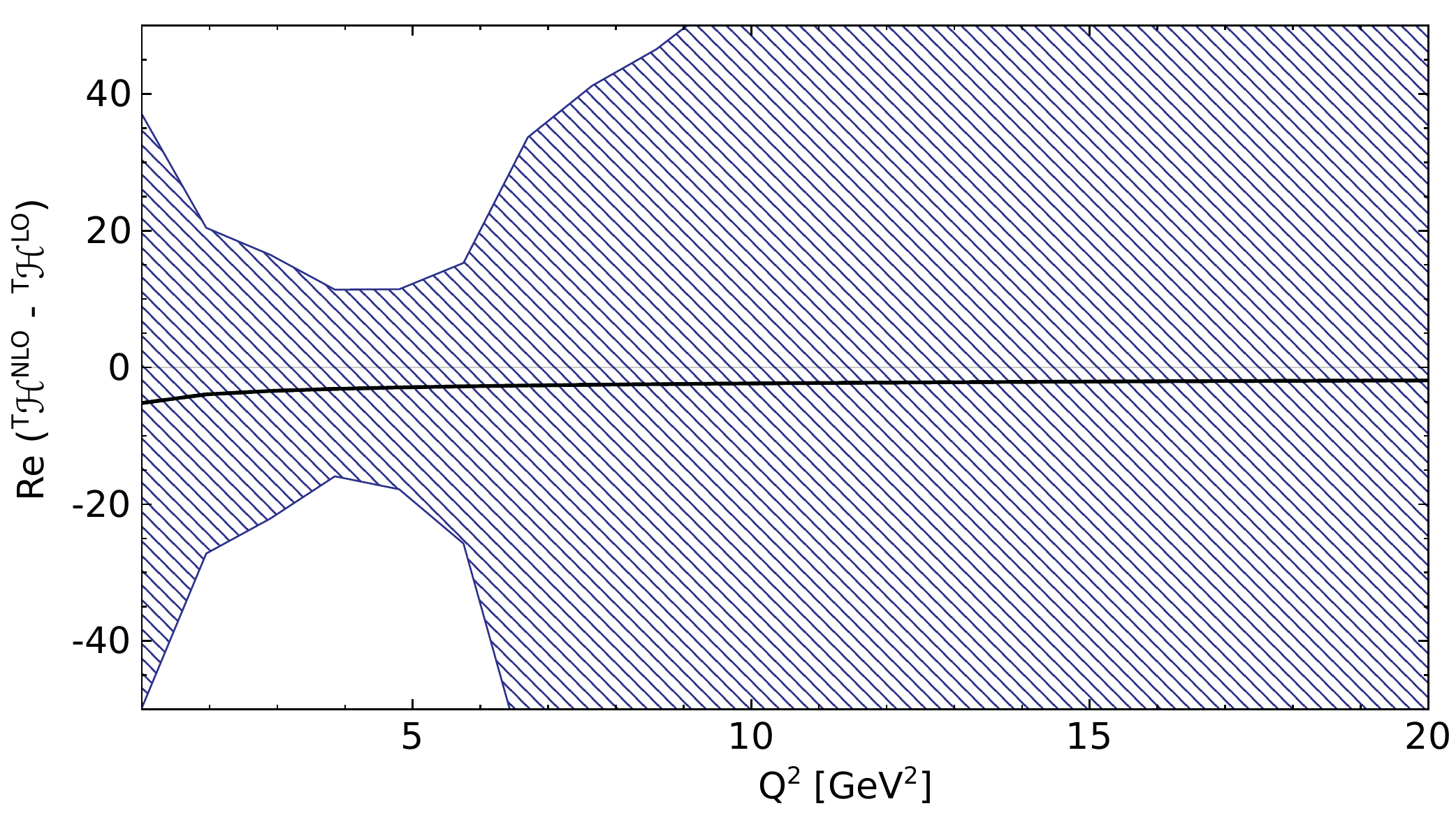}
\caption{Imaginary (up) and real (down) part of the NLO contribution to the spacelike-to-timelike relation for CFF $\mathcal{H}$ for $\xi = 0.1$ and $t=-0.1$ GeV$^2$ as a function of $Q^2$. For the further description see the caption of Fig. \ref{fig:TCS_CFFs}.}
\label{fig:TCS_CFFs_OnlyNLO}
\end{center}
\end{figure}

For illustration we focus now on CFF ${\cal H}$. In Fig. \ref{fig:DVCS_CFFs}  we  show the extracted DVCS CFF $^S\cal{H}$ (shaded gray band) as a function of $\xi$ for exemplary kinematics of $Q^2 =2$~GeV$^2$, $t = -0.3$ GeV$^2$. For comparison, we also present a model prediction based on the Goloskokov-Kroll (GK) parametrization of GPDs \cite{Goloskokov:2005sd, Goloskokov:2007nt, Goloskokov:2009ia}, obtained with LO (dashed line) and NLO (solid line) coefficient functions. All those quantities are used to perform predictions for TCS CFF $^T\cal{H}$, which are presented in Fig.~\ref{fig:TCS_CFFs}. The bigger uncertainty of the NLO result (dashed blue band) as compared to the LO one (shaded red band), reflects the fact that the available data do not constrain much the $Q^2$ dependence of DVCS CFFs, \cf Eq. \eqref{eq:cffH2cffTH}. To illustrate that, we present in Fig.~\ref{fig:TCS_CFFs_OnlyNLO} the difference between LO and NLO results, \ie the second term of Eq. (\ref{eq:cffH2cffTH}), as a function of $Q^2$. The solid line represents the GK model predictions, with very mild $Q^2$ dependence. Although in this model only the forward evolution is implemented, we have checked that the result obtained with the full evolution equations (treating GK as an input at $\mu_F =2$~GeV) is similar. The prediction based on the unbiased fit to the DVCS data (represented by the dashed blue band) has a large uncertainty, reflecting the sparseness and limited range in $Q^2$ of the used data. A future electron-ion collider \cite{Accardi:2012qut, Aschenauer:2017jsk}  will offer a much needed large lever arm in $Q^2$. We also observe a similar behaviour  in the remaining CFFs, manifested most dramatically in the case of poorly known CFF $\widetilde{\cal{E}}$.

This big uncertainty on the $Q^2$ dependence of the NLO prediction of TCS CFFs allows to draw some important conclusions. Firstly, one can expect very strong impact of the near future TCS measurements on the extraction of CFFs, and hence of GPDs. Secondly, it shows the necessity of including  the NLO effects in phenomenological studies of the DVCS and TCS data even in the so called ``valence region''. 
 
%%%%%%%%%%%%%%%%%%%%%%%%
%%%%%%%%%%%%%%%%%%%%%%%%
%%%%%%%%%%%%%%%%%%%%%%%%
%%%%%%%%%%%%%%%%%%%%%%%%
\section{TCS observables}
\label{sec:TCS_observables}
We will now focus on data-driven predictions for  observables in photoproduction of a lepton pair:
\begin{equation}
\label{eq:tcs-kin}
 \gamma(q)~N(p_1) \to l^-(k)~l^+(k')~N'(p_2)\,.   
\end{equation}
Similarly to the case of DVCS, the electromagnetic process referred to as Bethe-Heitler (BH) interferes with TCS at the level of amplitudes. In BH the lepton pair is radiatively generated by the incoming photon in bremsstrahlung, while in TCS it comes from the conversion of the virtual photon emitted by partons. The cross-section for 
photoproduction of a lepton pair can be expressed in the following way:
\begin{gather}
\frac{d \sigma}{dQ'^2~dt~d\phi~d\cos\theta} = 
\frac{d \sigma_{\mathrm{BH}}}{dQ'^2~dt~d\phi~d\cos\theta} \nonumber \\
+ \frac{d \sigma_{\mathrm{TCS}}}{dQ'^2~dt~d\phi~d\cos\theta}+
\frac{d \sigma_{\mathrm{INT}}}{dQ'^2~dt~d\phi~d\cos\theta} \,,
\end{gather}
where one can recognize contributions coming from BH, TCS and their interference. The angles $\theta$ and $\phi$ are the polar and azimuthal angles of $\vec{k}$ in the lepton-pair rest frame, respectively, with reference to a coordinate system with $z$-axis along $-\vec{p_2}$. 

\subsection{Unpolarized cross section}

\begin{figure}[!h]   
\includegraphics[width=\figWidth]{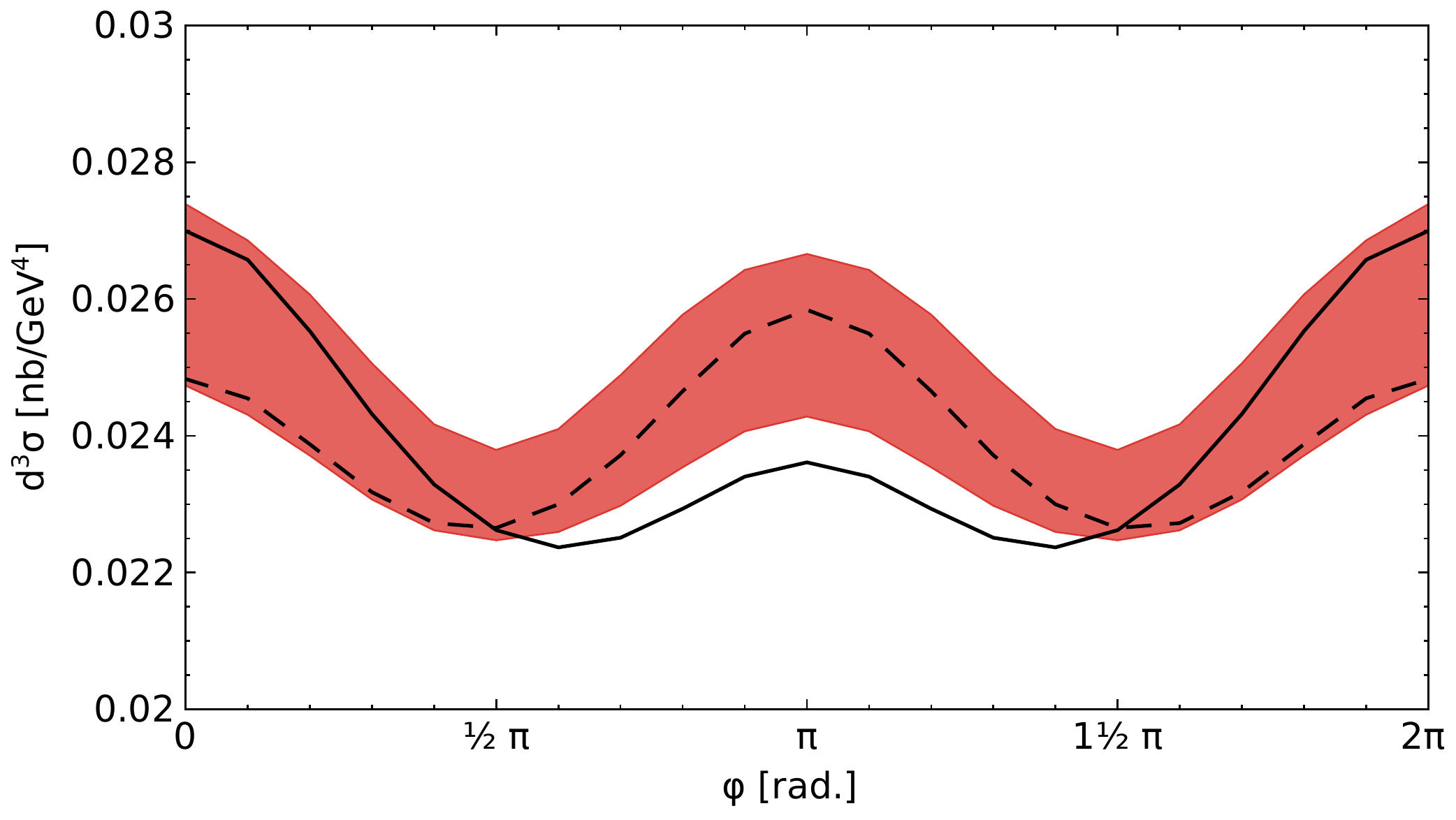}
\includegraphics[width=\figWidth]{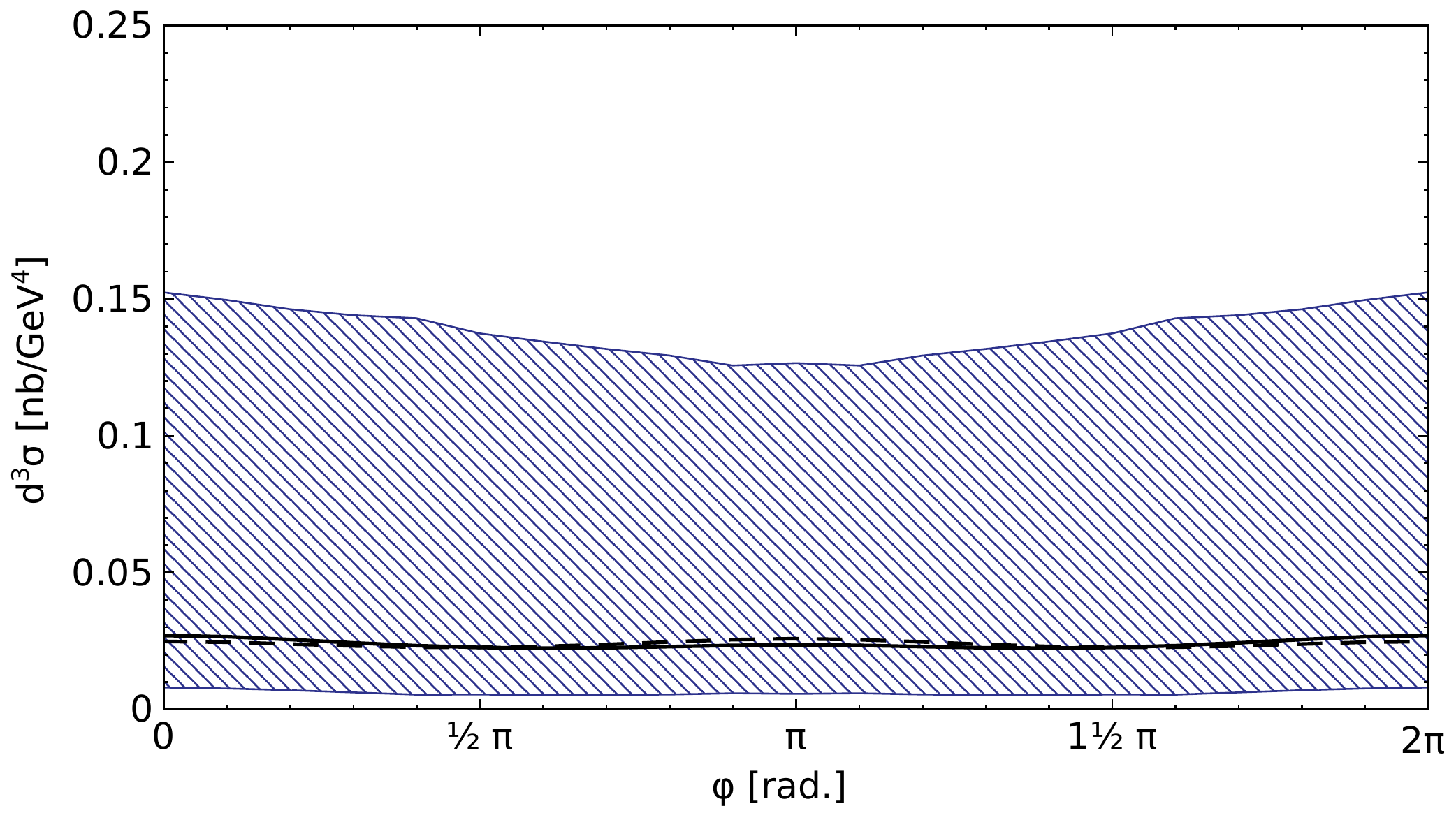}
\caption{Differential TCS cross section integrated over $\theta\in (\pi/4,3\pi/4)$ for $Q'^2 = 4$ GeV$^2$, $t=-0.1$ GeV$^2$ and the photon beam energy $E_\gamma = 10$ GeV as a function of the angle $\phi$. In the upper (lower) panel the data-driven predictions evaluated using LO (NLO) spacelike-to-timelike relations are shown. The dashed (solid) lines correspond to the GK GPD model \cite{Goloskokov:2005sd, Goloskokov:2007nt, Goloskokov:2009ia} evaluated with LO (NLO) TCS coefficient functions (the curves are the same in both panels). Note the different scales for the upper and lower panels. For the further description see the caption of Fig. \ref{fig:TCS_CFFs}.
}
\label{fig:cs_LO}
\end{figure}

The BH process dominates over TCS, especially in the moderate photon energy range \cite{Pire:2008ea}  and for the angle $\theta$ close to either 0 or $\pi$. To get a better sensitivity to the TCS signal, coming mainly from the interference term, we are integrating the cross section over $\theta$ between $\pi/4$ and $3\pi/4$. The LO prediction for the differential cross section as a function of the angle $\phi$ for $Q'^2 = 4$ GeV$^2$, $t=-0.1$ GeV$^2$ and the energy of the photon beam $E_\gamma = 10$ GeV is presented in the upper panel of Fig.~\ref{fig:cs_LO}. In the lower panel of that figure the NLO prediction is shown, and as in the case of the Fig.~\ref{fig:TCS_CFFs} we see a big uncertainty reflecting our limited knowledge of DVCS CFFs, especially of their $Q^2$ dependence. This demonstrates the big potential of TCS measurements.

\subsection{R ratio}
\begin{figure}[!h]
\begin{center}
\includegraphics[width=\figWidth]{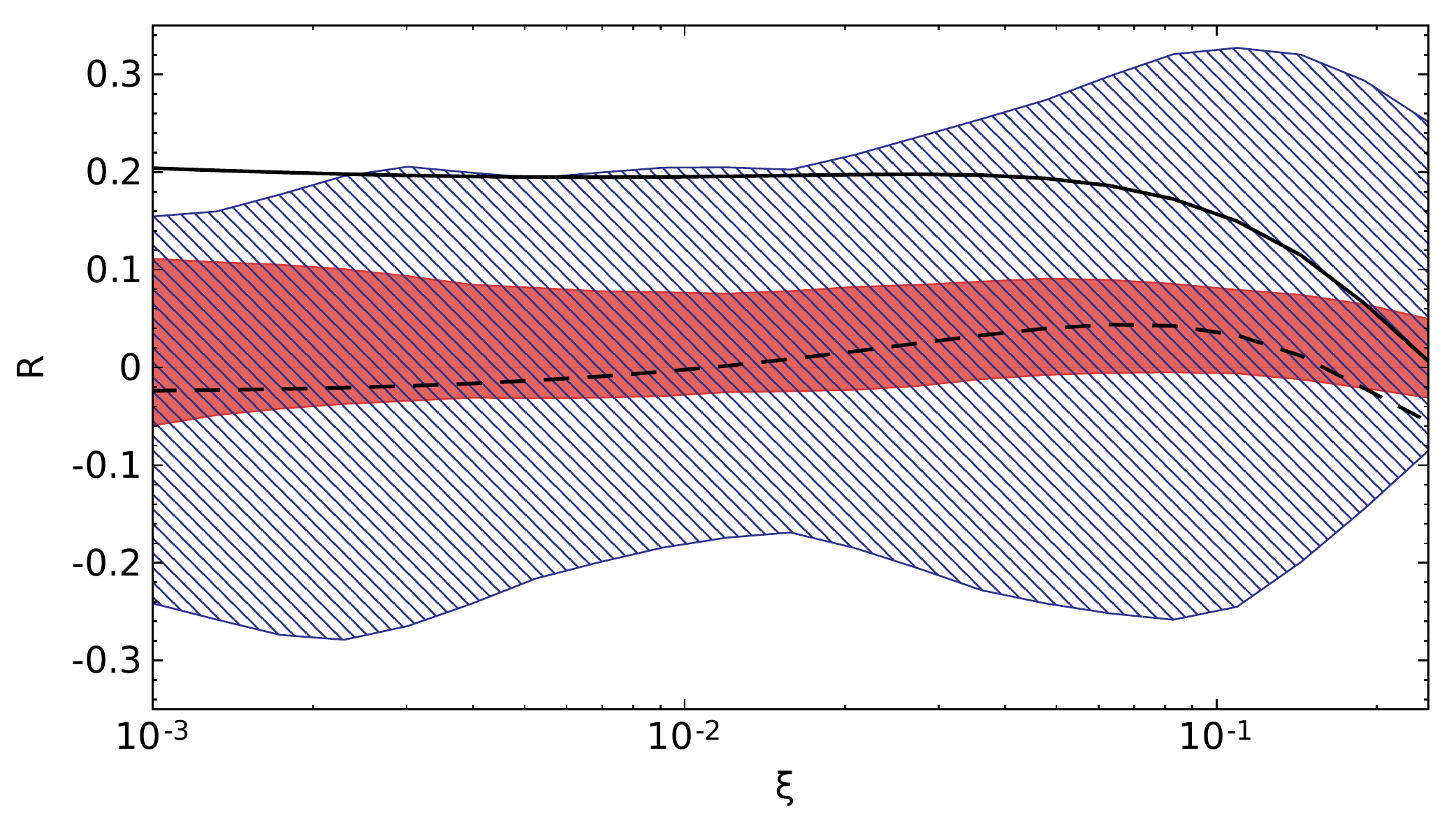}
\caption{Ratio $R$ defined in Eq. \eqref{eq:R} evaluated with LO and NLO spacelike-to-timelike relations for $Q'^2 = 4$ GeV$^2$, $t=-0.35$~GeV$^2$ as a function of $\xi$. For the further description see the caption of Fig. \ref{fig:TCS_CFFs}.
}
\label{fig:R}
\end{center}
\end{figure}

An important observable in the phenomenology of TCS is the  $R$ ratio, introduced in Ref. \cite{Berger:2001xd}:
\begin{equation}
R = \frac{
\displaystyle
2\int_0^{2\pi}\cos \phi ~d\phi \int_{\pi/4}^{3\pi/4}d \theta \frac{dS}{dQ'^2 dt d\phi d\theta}
}{
\displaystyle
\int_0^{2\pi} ~d\phi \int_{\pi/4}^{3\pi/4}d \theta \frac{dS}{dQ'^2 dt d\phi d\theta}
} \,,
\label{eq:R}
\end{equation}
where $S$ is the weighted cross section \cite{Berger:2001xd}:
\begin{equation}
 \frac{dS}{dQ'^2 dt d\phi d\theta} = \frac{L(\theta,\phi)}{L_0(\theta)} \frac{d\sigma}{dQ'^2 dt d\phi d\theta}\,,
\label{eq:dS}
\end{equation}
and where $L=(q-k)\cdot (q-k')$ and $L_0=L(t\to 0,M^2\to 0)=Q'^4 \sin^2\theta/4$ with the notations of Eq.~(\ref{eq:tcs-kin}). This observable is particularly interesting, as due to a different charge conjugation properties of the lepton pair in the BH and TCS processes, it projects out the interference term that is linear in CFFs. Moreover, this observable has a special sensitivity to the real part of CFF $^T\cal{H}$. Our prediction for the ratio $R$ as a function of $\xi$ is shown in Fig. \ref{fig:R} for the values of $Q'^2 = 4$~GeV$^2$ and $t=-0.35$ GeV$^2$. As noted in Ref.~\cite{Moutarde:2013qs}, the model predictions with LO and NLO coefficient functions (denoted in Fig.~\ref{fig:R}  by the dashed and solid lines, respectively) differ by a large factor\footnote{There is an error in numerical estimates of the ratio  $R$ in Ref.~\cite{Moutarde:2013qs}, where the BH contribution in the denominator was mistakenly multiplied by two.}.

\subsection{Circular asymmetry}
The circular polarization of photons can be generated in bremsstrahlung of longitudinally polarized leptons. The asymmetry probing various states of circular beam polarization is interesting as it singles out specific elements of the interference contribution to the cross section. This contribution reads \cite{Berger:2001xd}:
\begin{gather}
\frac{d \sigma_{\mathrm{INT}}}{dQ'^2\, dt\, d(\cos\theta)\,d\phi}  \nonumber \\  
=
\frac{d \sigma_{\mathrm{INT}}^{\mathrm{unpol}}}{dQ'^2\, dt\, d(\cos\theta)\,d\phi} + \frac{d \sigma_{\mathrm{INT}}^{\mathrm{circular}}}{dQ'^2\, dt\, d(\cos\theta)\,d\phi} \nonumber \\
= 
\frac{\alpha^3_{em}}{4\pi s^2}\, \frac{1}{-t}\, \frac{M}{Q'}\,
\frac{1}{\tau \sqrt{1-\tau}}\, \frac{L_0}{L}
\, \cos\phi\, \frac{1+\cos^2\theta}{\sin\theta}\,
    \mathrm{Re}\widetilde{M}^{--} \nonumber \\
 -
 \nu\; \frac{\alpha^3_{em}}{4\pi s^2}\, \frac{1}{-t}\, 
\frac{M}{Q'}\, \frac{1}{\tau \sqrt{1-\tau}}\, \frac{L_0}{L} 
 \sin\phi\,
\frac{1+\cos^2\theta}{\sin\theta}\, \mathrm{Im}\widetilde{M}^{--} 
\,,
\label{helasy}
\end{gather}
with
\begin{align}
\label{mmimi}
\widetilde{M}^{--} &= 
\frac{2\sqrt{t_0-t}}{M}\, \frac{1-\xi}{1+\xi}\,
\nonumber \\
&\times \left[ F_1 {\cal H} - \xi (F_1+F_2)\, \widetilde{\cal H} -
\frac{t}{4M^2} \, F_2\, {\cal E} \,\right] \,,
\end{align}
where $\alpha_{em}$ is the fine structure constant, $s$ is the energy squared calculated in the photon-proton CMS, and $F_1$, $F_2$ are the elastic form factors. The circular polarization state is given by $\nu = \pm 1$. Note, that in Eq. \eqref{helasy} the polarization dependent and independent parts are simply related by an exchange of $\sin \phi \leftrightarrow \cos \phi$ and $\mathrm{Im} \leftrightarrow \mathrm{Re}$. Imaginary parts of CFFs can be accessed through the asymmetry with respect to $\nu$:
\begin{equation}
A_{CU} = \frac{\sigma(\nu = +1)-\sigma(\nu = -1)}{\sigma(\nu = +1)+\sigma(\nu = -1)} \,.    
\end{equation}
The denominator of this asymmetry is dominated by the square of BH amplitude, which is almost flat in $\phi$. 

\begin{figure}[!h]
\begin{center}
\includegraphics[width=\figWidth]{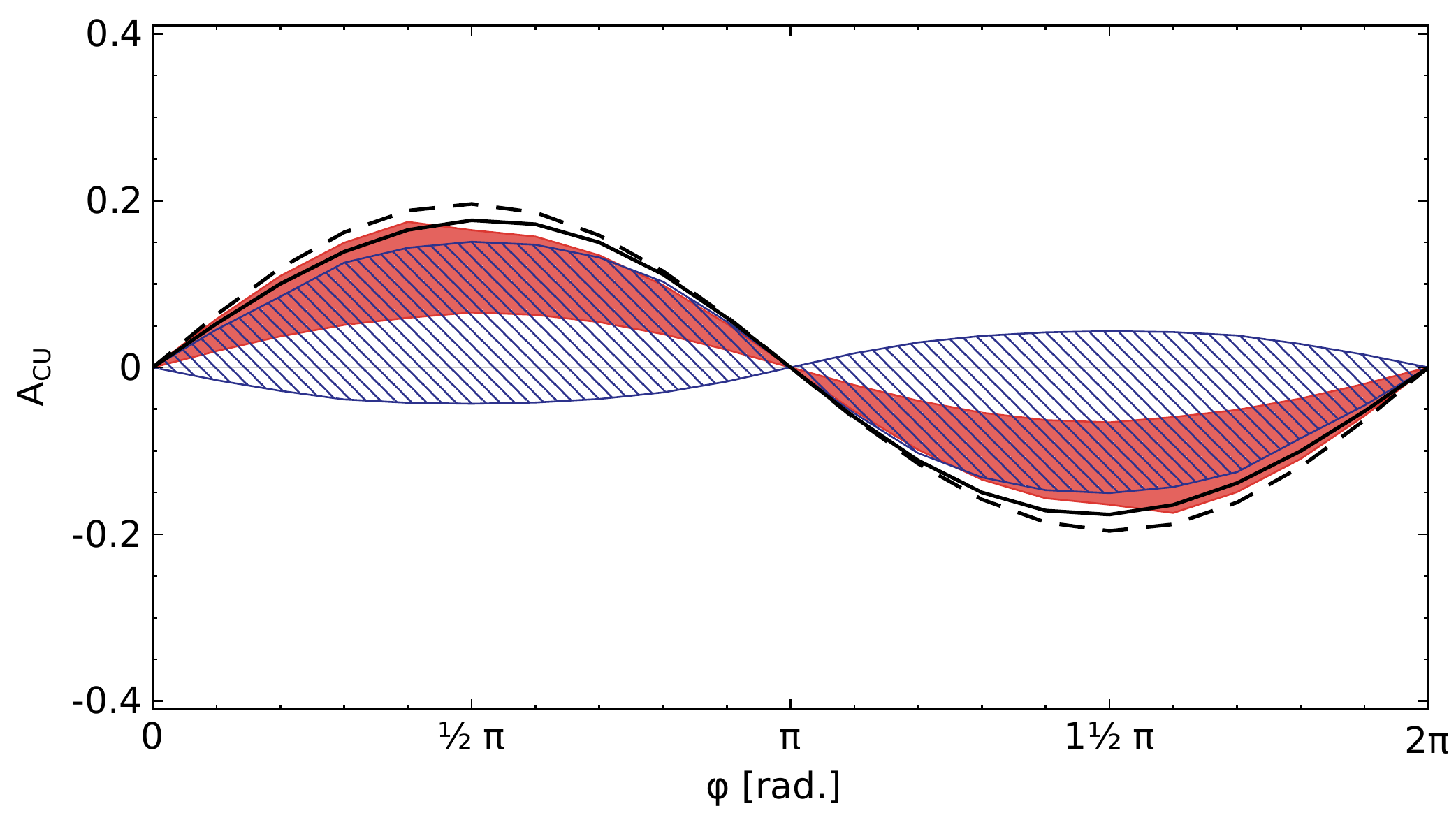}
\end{center}
\caption{Circular asymmetry $A_{CU}$ evaluated with LO and NLO spacelike-to-timelike relations for $Q'^2 = 4$ GeV$^2$, $t=-0.1$~GeV$^2$ and $E_\gamma =10$ GeV as a function of $\phi$.  The cross sections used to evaluate the asymmetry are integrated over $\theta\in (\pi/4,3\pi/4)$. For the further description see the caption of Fig. \ref{fig:TCS_CFFs}.
}
\label{fig:ACU_phi}
\end{figure}

\begin{figure}[!h]
\begin{center}
\includegraphics[width=\figWidth]{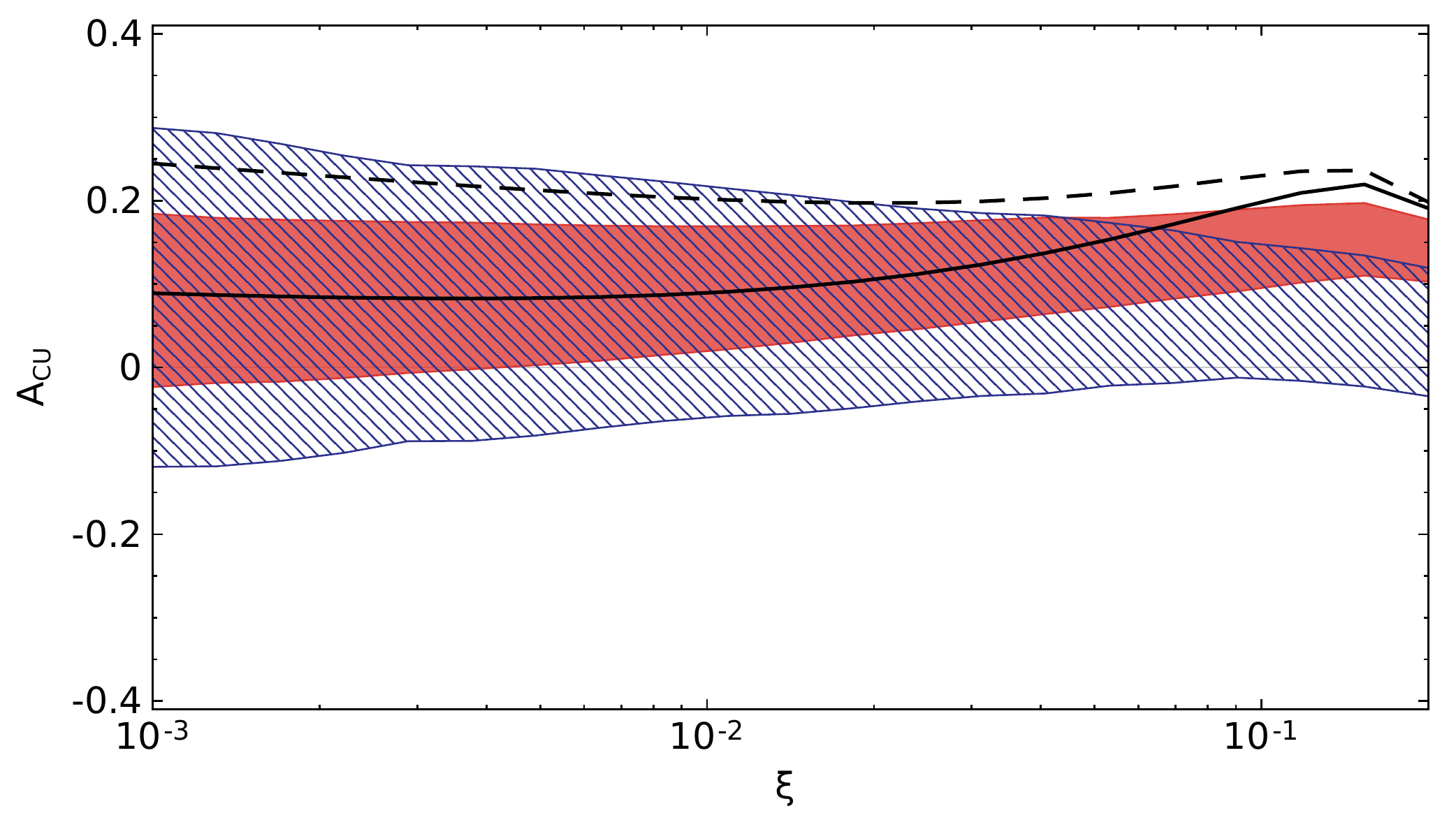}
\end{center}
\caption{Circular asymmetry $A_{CU}$ evaluated with LO and NLO spacelike-to-timelike relations for $Q'^2 = 4$ GeV$^2$, $t=-0.1$~GeV$^2$ and $\phi = \pi/2$ as a function of $\xi$. The cross sections used to evaluate the asymmetry are integrated over $\theta\in (\pi/4,3\pi/4)$. For the further description see the caption of Fig. \ref{fig:TCS_CFFs}.
}
\label{fig:ACU_xi}
\end{figure}

The prediction for $A_{CU}$ asymmetry as a function of $\phi$ is presented in Fig. \ref{fig:ACU_phi} for $Q'^2 = 4$ GeV$^2$ and $t=-0.1$ GeV$^2$ at LO and NLO. The magnitude of the asymmetry, \ie its value at $\phi = \pi/2$, is presented in Fig. \ref{fig:ACU_xi} as a function of $\xi$. In both plots the asymmetries evaluated from cross sections integrated over $\theta$ in the range $(\pi/4,3\pi/4)$ are shown. 
 
\subsection{Linear Polarization}

Experimental techniques recently developed at JLab enable the study of TCS with an intense beam of linearly polarized photons. We have shown in Ref. \cite{Goritschnig:2014eba} that observables based on the angular distribution of the lepton pair project out cross-section contributions associated with certain combination of GPDs, and in particular are sensitive to  poorly known polarized GPDs $\widetilde{H}$. 

\begin{figure}[!h]
\begin{center}
\includegraphics[width=\figWidth]{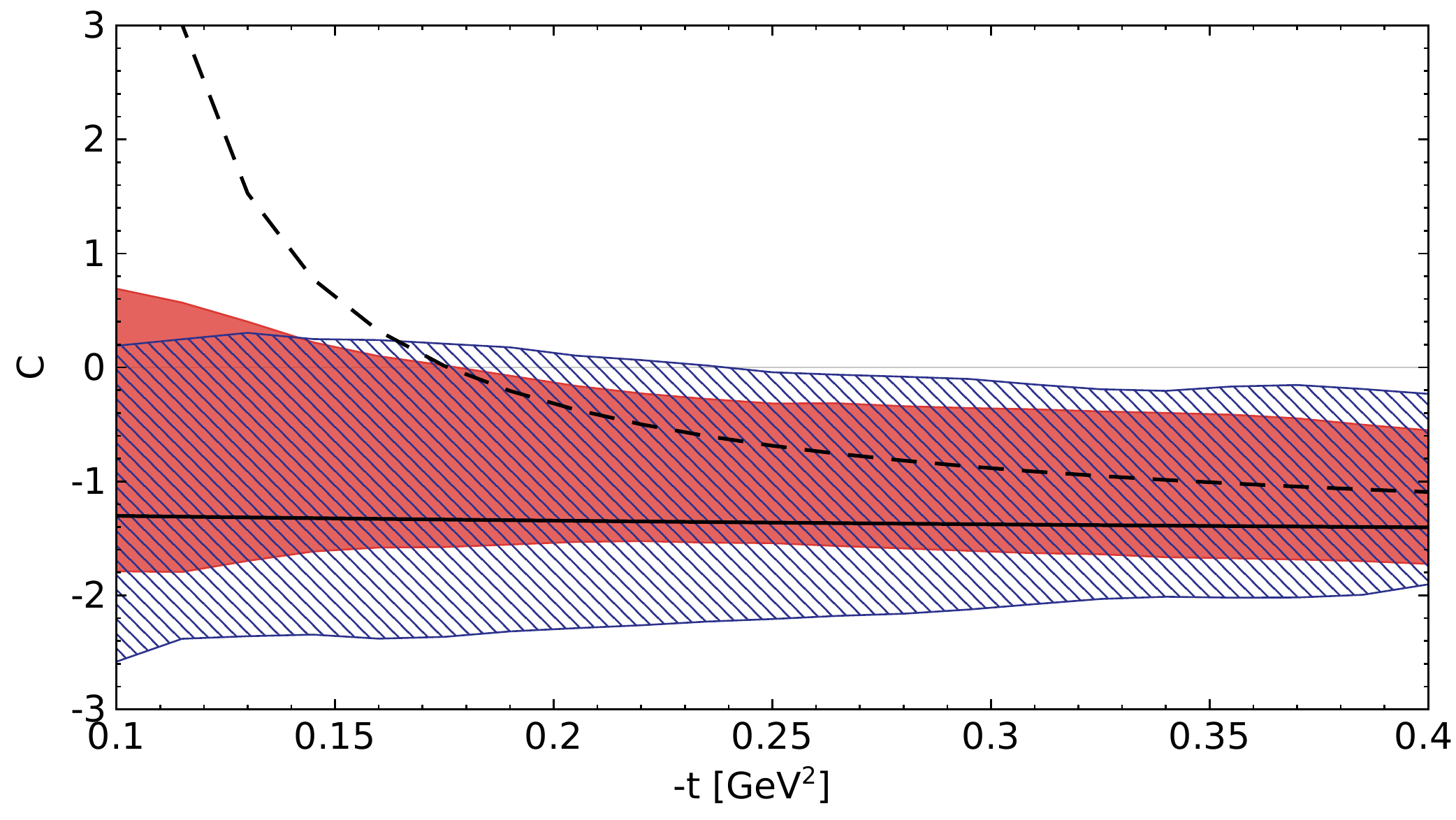}
\end{center}
\caption{Ratio $C$ defined in Eq. \eqref{eq:C} evaluated with LO and NLO spacelike-to-timelike relations for $Q'^2 = 4$ GeV$^2$, $\xi=0.1$ as a function of $t$. For the further description see the caption of Fig. \ref{fig:TCS_CFFs}.
}
\label{fig:C_t}
\end{figure}

In the description of TCS with linearly polarized photons, one needs to introduce an additional angle $\Phi_h$ between the polarization vector and the hadronic plane. The  contribution to interference cross section due to this polarization reads \cite{Goritschnig:2014eba}: 
\begin{gather}
\frac{d\sigma_{\mathrm{INT}}^{\mathrm{linpol}}}{dQ'^2 dt d(\cos\theta)d\phi d \Phi_h} 
 = \nonumber \\
 -  
\frac{\alpha_{em}^3}{16\pi^2s^2}\frac{1}{Q'^2} \left(\frac{4s\mid{\Delta}_\perp\mid}{Q't}\right) \,  
           \bigg(\sin\theta\cos(2\Phi_h + 3\phi) \bigg) \nonumber \\
\times
\text{Re}\left[\mathcal{H}F_1 - \frac{t}{4M^2} \mathcal{E}F_2 + \xi\widetilde{\mathcal{H}}(F_1 + F_2)) \right] \,,
\label{eq:sigmaX-INT}
\end{gather}
where $\Delta_\perp$ is the transverse momentum transfer on the proton target. This contribution allows us to define the following observable, which is sensitive only to the interference term and provides us with information about CFF $\widetilde{\mathcal{H}}$:
\begin{gather}
C = 
\nonumber \\
\frac{
\int_0^{2\pi} d\Phi_h 2\int_0^{2\pi} d\phi\cos(\phi) \int_{\pi/4}^{3\pi/4}\sin \theta d \theta d^{5} \sigma
}{
2\int_0^{2\pi} d\Phi_h \cos(2\Phi_h) 2\int_0^{2\pi} d\phi\cos(3\phi) \int_{\pi/4}^{3\pi/4}\sin \theta d \theta d^{5} \sigma
} 
= \nonumber \\
\frac{2-3\pi}{2+\pi}
\frac
{
	 \text{Re}\left[\mathcal{H}F_1 - \frac{t}{4M^2} \mathcal{E}F_2 - 	\xi\widetilde{\mathcal{H}}(F_1 + F_2)\right]
}
{
	 \text{Re}\left[\mathcal{H}F_1 - \frac{t}{4M^2} \mathcal{E}F_2 + 	\xi\widetilde{\mathcal{H}}(F_1 + F_2)\right]
}\,.
\label{eq:C}
\end{gather}
We present LO and NLO data-driven predictions of that observable in Fig. \ref{fig:C_t}. It should be noted that the difference between LO and NLO prediction uncertainties is smaller than in previous cases, but these uncertainties are quite sizable anyway, which makes their measurement particularly important for the determination of GPD $\widetilde {H}$.

\subsection{Transverse target asymmetry}

\begin{figure}[!h]
\begin{center}
\includegraphics[width=\figWidth]{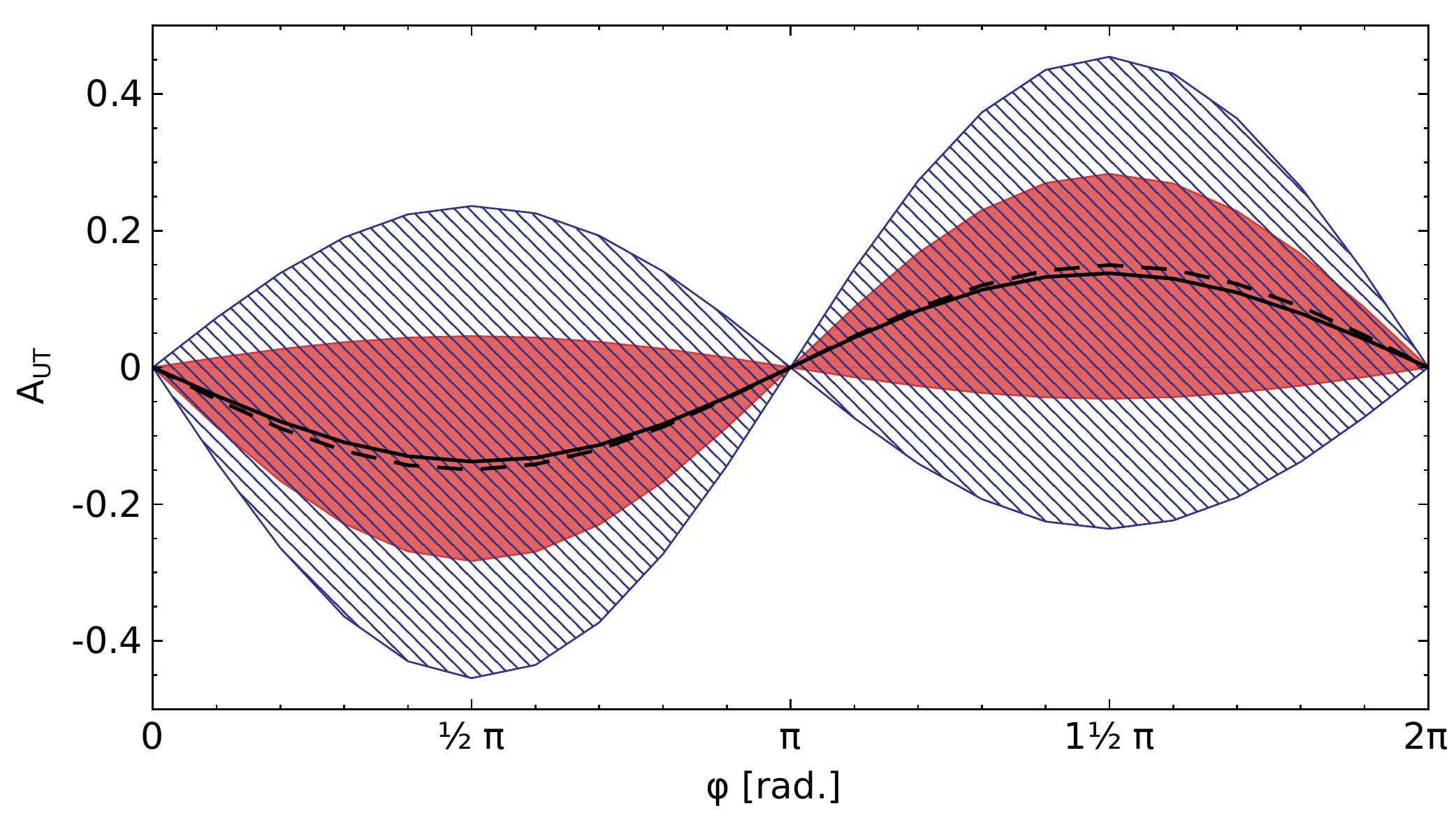}
\end{center}
\caption{Transverse target spin asymmetry $A_{UT}$ evaluated with LO and NLO spacelike-to-timelike relations for $Q'^2 = 4$ GeV$^2$, $t=t_0$ and $E_\gamma = 10$ GeV as a function of $\varphi_S$. The cross sections used to evaluate the asymmetry are integrated over $\theta\in (\pi/4,3\pi/4)$. For the further description see the caption of Fig. \ref{fig:TCS_CFFs}.
}
\label{fig:AUT_phi}
\end{figure}

Finally, we present our new results for the transverse target spin asymmetry. In the limit of vanishing transverse momentum transfer, \ie for $t = t_0$, the only part of the cross section that depends on the transverse target spin polarization comes from the interference term and it reads:
\begin{gather}    
    \frac{d\sigma_{\mathrm{INT}}^{\mathrm{tpol}}}{dQ'^2 d(\cos\theta) d \phi dt d\varphi_S} = \nonumber \\
    -\frac{\alpha_{em}^3}{16\pi^2}\frac{M}{s^2 t Q'}(F_1+F_2) \sin\theta \sin \varphi_S\xi
    \times \Bigg{(}  \mathcal{A}\Big{[} \Im(\mathcal{H})  \nonumber \\
    - \frac{\xi^2}{1-\xi^2}\Im(\mathcal{E})\Big{]} - \mathcal{B} \Big{[} \Im(\widetilde{\mathcal{H}}) +\frac{t}{4M^2}\Im(\widetilde{\mathcal{E}}) \Big{]} \Bigg{)} ,
\end{gather}
where 
\begin{equation}
\mathcal{A} = 4 \Big{(} 1 + \frac{2Q'^2 (Q'^2-t)\cos^2\theta}{(Q'^2 - t)^2 - b^2} \Big{)}
\end{equation} 
and 
\begin{equation}
\mathcal{B} = 2 \Big{[} \frac{t-b+Q'^2}{t-b-Q'^2} + \frac{t+b+Q'^2}{t+b-Q'^2} \Big{]} \,.
\end{equation}
Here, $b = 2 (k-k')\cdot(p-p')$ and $\varphi_S$ is the angle between the leptonic plane and the nucleon spin direction. In the limit of $|t|\ll s,Q'^2$, this result simplifies to:
\begin{gather}   
    \frac{d\sigma_{\mathrm{INT}}^{\mathrm{tpol}}}{dQ'^2 d(\cos\theta)  d \phi dt d\varphi_S} = \nonumber \\
    -\frac{\alpha_{em}^3}{4\pi^2}\frac{M}{s^2 t Q'}(F_1+F_2) \frac{1+\cos^2\theta}{\sin\theta} \sin\varphi_S ~ \xi \nonumber \\ \times\Im \Big{[} 
    \mathcal{H} - \frac{\xi^2}{1-\xi^2}\mathcal{E} +  \widetilde{\mathcal{H}} +\frac{t}{4M^2}\widetilde{\mathcal{E}}\Big{]}.
\end{gather}
The transverse spin asymmetry:
\begin{equation}
    A_{UT}(\varphi_S)=\frac{\sigma(\varphi_S) - \sigma(\varphi_S-\pi)}{\sigma(\varphi_S) + \sigma(\varphi_S-\pi)}\,,
\end{equation}
at the point $Q'^2 = 4$ GeV$^2$, $t=t_0$ and $E_\gamma =10$ GeV in presented in Fig. \ref{fig:AUT_phi} as a function of $\varphi_S$. Model predictions based on the GK GPDs (with LO and NLO coefficient functions) estimate the size of $A_{UT}$ of the order of $15\%$. Data-driven predictions allow for a sizeable asymmetry as well.

\section{Summary}
We presented the first multi-channel data-driven analysis of exclusive processes relying on a global fit of CFFs. Among impact studies of GPD-related channels, the present one is also the first going beyond the LO approximation, providing systematic comparisons of predictions obtained with coefficient functions evaluated at LO or NLO. Our analysis is characterized by a low model-dependency, as essentially it is done at the level of DVCS and TCS amplitudes parameterized with neural networks, and we only use spacelike-to-timelike relations to connect those two reactions.

Our data-driven study of the timelike Compton scattering process has demonstrated the crucial need of lepton pair photoproduction data to access in a sensible way GPDs of the nucleon. It also showed in a quantitative way why any extraction of GPDs  based on a leading order analysis of DVCS experimental data is very incomplete and then very model-dependent.  In particular, aiming at a reasonable understanding of the $Q^2$ dependence of the GPDs depends much on a concomitant analysis of the DVCS and TCS reactions. 

We did not discuss the important issue of the needed twist-3 contributions to preserve QED gauge invariance, of the finite-$t$ and target mass corrections, nor their phenomenological consequences \cite{Anikin:2000em,Belitsky:2000vx,Belitsky:2001yp,Braun:2012hq,Braun:2014sta}, but we acknowledge that these required refinements will definitely be needed for a state of the art extraction of CFFs and GPDs when much more data will be available. Although we did not address explicitly the neutron \cite{Boer:2015cwa} or the light nucleus \cite{Berger:2001zb, Kirchner:2003wt,Scopetta:2004kj, Dupre:2015jha} target case where the DVCS extraction of CFFs has not yet been performed through our ANN technique, we believe that the same conclusions apply as well to these cases.

\begin{acknowledgements}
This project was supported by the European Union's Horizon 2020 research and innovation programme under grant agreement No 824093 and by the Grant No. 2017/26/M/ST2/01074 of the National Science Centre, Poland. The project is co-financed by the Polish National Agency for Academic Exchange and by the COPIN-IN2P3 Agreement.
\end{acknowledgements}

\bibliographystyle{spphys}
\bibliography{bibliography}

\end{document}